# Solution of the nonstationary diffusion equation for interstitial impurity atoms by the method of Green's functions

O.I. Velichko* and O.N. Burunova

Department of Physics, Belarusian State University of Informatics and Radioelectronics, 6, P.Brovki Str., Minsk, 220013 Belarus, E-mail: oleg_velichko@lycos.com

**Abstract.** On the basis of the Green's function method, analytical solutions of the diffusion equation which describes nonstationary migration of nonequilibrium interstitial impurity atoms have been derived. It is supposed that the initial distribution of nonequilibrium impurity interstitials is formed due to ion implantation and, therefore, is described by the Gaussian function. The condition of the constant concentration of impurity interstitials (the Dirichlet boundary condition) or reflecting boundary condition was imposed on the surface of a semiconductor. The Dirichlet boundary condition was also enforced for the concentration of impurity interstitials in the infinity, i.e., in the bulk of a semiconductor. On the basis of the solutions derived the redistribution of ion-implanted boron in silicon substrate during low-temperature thermal treatment has been simulated. The calculated profile of boron atoms after annealing agrees well with experimental data. It means that the analytical solutions derived can be used both for verifying the numerical results and for modeling the long-range migration of nonequilibrium impurity interstitials during low-temperature thermal treatments.

# Введение

Для создания современных полупроводниковых приборов и интегральных микросхем широко используются методы легирования посредством ионной имплантации в сочетании с последующей термической обработкой полупроводниковых слоев. Схематическое изображение этого способа легирования приведено на Рис.1.

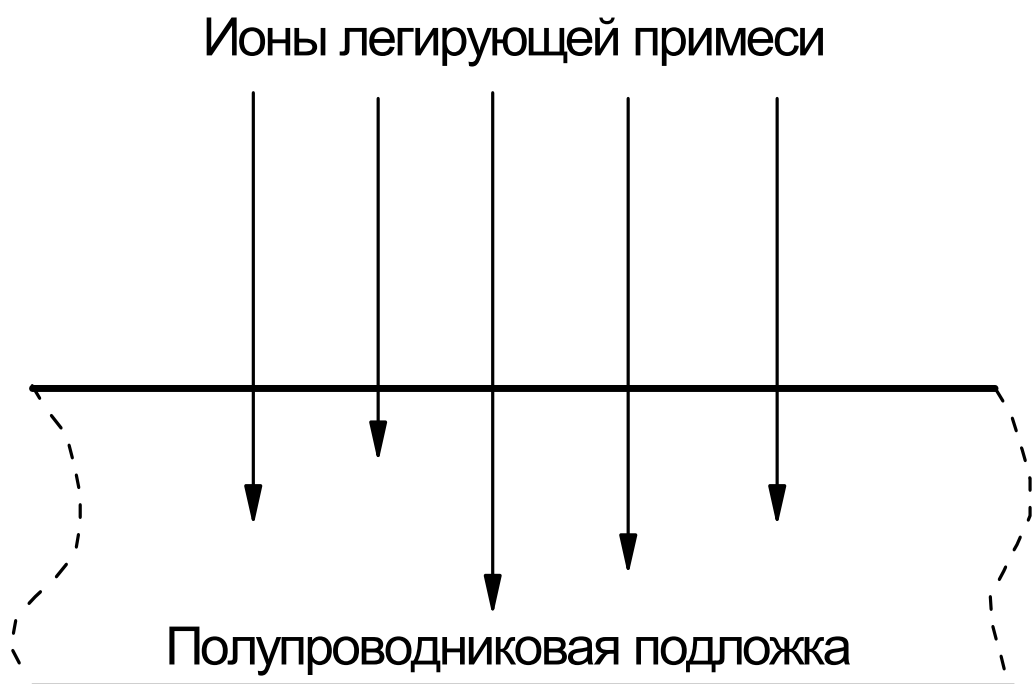

Рис.1. Легирование посредством ионной имплантации

В процессе ионного внедрения происходит генерация большого количества неравновесных постимплантационных дефектов, в том числе вакансий, собственных межузельных атомов, а также межузельных атомов примеси. Для удаления постимплантационных дефектов и электрической активации примеси применяется термическая обработка при температурах ~700 $^o$C и выше. В процессе термообработки происходит диффузионное перераспределение примесных атомов. Для уменьшения этого перераспределения и улучшения электрофизических параметров приборов широко применяются быстрые [1-3] или даже пиковые (spike annealing) высокотемпературные отжиги [1,4], либо короткие низкотемпературные обработки [5,6]. Как следует из экспериментальных данных, в этих случаях может наблюдаться длиннопробежная миграция неравновесных межузельных атомов примеси, что проявляется в формировании характерных протяженных "хвостов" в области низкой концентрации примесных атомов [1,5,6]. Типичный профиль распределения концентрации атомов ионно-имплантированного бора после низкотемпературного термического отжига, полученный в работе [5], представлен на Рис.2. Для экспериментов использовались образцы кристаллического кремния с удельным сопротивлением 16500 Ом см, которые имплантировались ионами бора. Энергия имплантации 70 кэВ, доза $10^{15}$ см$^{-2}$. Отжиг проводился в атмосфере азота при температуре 800 $^o$C в течение 35 минут. Как видно из Рис.2, даже при такой низкой температуре отжига, имеет место существенная диффузия атомов бора в области с концентрацией меньшей $5 \times 10^6$ мкм$^{-3}$. Существующие программы моделирования процессов легирования полупроводников (см. например, [7,8]) не описывают этого явления.

В работе [9] было предположено, что образование протяженных "хвостов" в области низкой концентрации примесных атомов, которое имеет место при перераспределении ионно-имплантированной примеси в случае коротких низкотемпературных термообработок, есть следствие длиннопробежной миграции неравновесных межузельных атомов примеси. Такие неравновесные межузельные атомы примеси могут образовываться в ионно-имплантированных слоя в результате того, что часть примесных атомов оказывается в межузельном положении непосредственно после имплантации ионов, либо переходит в межузельное положение вследствие распада части имплантационных дефектов на начальной стадии отжига. В случае, когда можно пренебречь процессами кластерообразования и преципитации атомов примеси, система уравнений, предложенная в [9] для описания процесса миграции неравновесных межузельных атомов примеси и их последующего перехода в положение замещения, имеет следующий вид:

**1. Закон сохранения атомов примеси в положении замещения**

$$\frac{\partial C(x,t)}{\partial t} = \frac{C^{AI}(x,t)}{\tau^{AI}} \, , \qquad (1)$$

**2. Уравнение диффузии неравновесных межузельных атомов примеси**

$$\frac{\partial C^{AI}}{\partial t} = d^{AI}\frac{\partial^2 C^{AI}}{\partial x^2} - \frac{C^{AI}}{\tau^{AI}} + G^{AIR}(x,t) \,, \qquad (2)$$

где

$C$ — концентрация атомов примеси в положении замещения;

$C^{AI}$ — концентрация неравновесных межузельных атомов примеси (МАП);

$d^{AI}$ — коэффициент диффузии этих атомов;

$\tau^{AI}$ — среднее время жизни неравновесного межузельного атома примеси;

$G^{AIR}$ — скорость генерации неравновесных межузельных атомов примеси в единице объема полупроводника в результате отжига имплантационных дефектов.

В работе [9] предполагалось, что переход неравновесных межузельных атомов примеси в положение замещения осуществляется в результате их рекомбинации с вакансиями, причем коэффициент рекомбинации $k^{AI}$ имеет следующий вид

$$k^{AI} = \frac{1}{\tau^{AI}} = k^{AIV} C^V \,, \qquad (3)$$

где

$k^{AIV}$ — постоянная рекомбинации;

$C^V$ — концентрация вакансий.

При описании процесса генерации неравновесных межузельных атомов примеси предполагалось, что

$$G^{AIR}(x,t) = f^R(x) g^R(t) = f^R(x)\exp\left(-\frac{t}{\tau^R}\right) \,, \qquad (4)$$

где $\tau^R$ — среднее время жизни неравновесных имплантационных дефектов, включающих атомы примеси.

Предложенную систему уравнений (1), (2) можно решить, используя численные методы. Необходимо отметить, что при более детальном рассмотрении диффузионных процессов, например, в случае необходимости учета процессов кластерообразования примесных атомов при высоких концентрациях примеси, рассмотрении различных зарядовых состояний точечных дефектов, уравнения (1) и (2) могут быть существенно усложнены. В частности, при учете зарядовых состояний вакансий и межузельных атомов примеси коэффициенты уравнения (2) представляют собой нелинейные функции концентрации носителей заряда в легированном слое [10]. Это означает, что очень важной является проблема достоверности полученного численного решения. Один из лучших способов проверить правильность приближенного численного решения — сравнить его с точным аналитическим решением.

Концентрация атомов бора, мкм⁻³

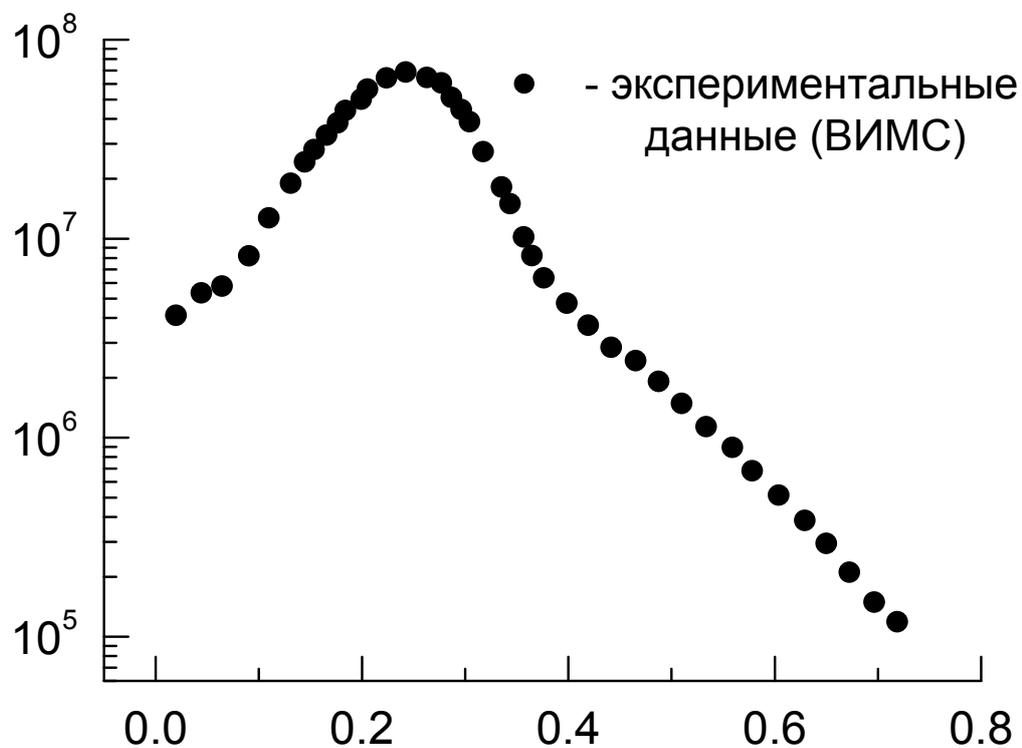

Рис.2. Профиль распределения концентрации атомов ионно-имплантированного бора после термического отжига в течение 35 минут при температуре 800 °C. Энергия имплантации 70 кэВ, доза $10^{15}$ см⁻² .

## ЦЕЛЬ ИССЛЕДОВАНИЯ

**Проведенный анализ позволяет сформулировать цель данного исследования:**

**1) Получить аналитическое решение нестационарного уравнения диффузии неравновесных межузельных атомов примеси, используя метод функций Грина и пакет "Математика 5.0", который обеспечивает проведение различных символьных вычислений, в том числе аналитическое интегрирование выражений, получаемых с помощью метода функций Грина.**

**2) Используя полученные решения, провести моделирования ряда диффузионных процессов и сравнить результаты моделирования с экспериментальными данными.**

## 1. Постановка задачи

Получим аналитическое решение уравнения диффузии неравновесных межузельных атомов примеси (2) на полубесконечной прямой $[0, +\infty]$. В качестве граничных условий используем вначале условия Дирихле (граничные условия первого рода) на обоих концах интервала

$$C^{AI}(0, t) = C_S^{AI} \ , \qquad\qquad C^{AI}(+\infty, t) = 0 \ , \qquad\qquad (5)$$

где $C_S^{AI}$ — концентрация межузельных атомов примеси на поверхности полупроводника.

Предположим, что отжиг имплантационных дефектов, содержащих атомы примеси имеет место только в течение очень короткой начальной стадии термообработки, причем $\tau^R \ll \tau^{AI}$. Тогда член $G^{AIR}(x,t)$ в правой части Ур.(2) можно считать равным нулю. Наличие большого количества освободившихся межузельных атомов примеси учтем в начальном условии для уравнения диффузии .(2). С той целью зададим начальное условие в виде

$$C^{AI}(x, 0) = C_0^{AI}(x) \ , \qquad\qquad (6)$$

где $C_0^{AI}(x)$ — распределение межузельных атомов примеси после имплантации (более точно после имплантации и отжига имплантационных дефектов, содержащих атомы примеси).

В данной работе предполагаем, что распределение неравновесных межузельных атомов примеси после имплантации и отжига имплантационных дефектов, содержащих атомы примеси, пропорционально распределению имплантированных атомов, которое можно приближенно описать функцией распределения Гаусса. Тогда, начальное условие (6) можно представить в следующем виде

$$C^{AI}(x, 0) = C_m^{AI} \exp\left[-\frac{(x - R_p)^2}{2\Delta R_p^2}\right], \qquad\qquad (7)$$

где

$C_m^{AI}$ — максимальная концентрация межузельных атомов примеси (МАП);

$R_p$ — средний проективный пробег ионов;

$\Delta R_p$ — страгглинг проективного пробега ионов.

Учитывая, что в рассматриваемом случае $G^{AIR}(x,t)=0$, приведем уравнение диффузии (2) к более удобному виду. С этой целью умножим обе части уравнения (2) на $\tau^{AI}$ и получим

$$\tau^{AI}\frac{\partial C^{AI}}{\partial t}=\tau^{AI} d^{AI}\frac{\partial^2 C^{AI}}{\partial x^2}-C^{AI}. \tag{8}$$

Введём следующие обозначения:

$l_{AI}=\sqrt{d^{AI}\tau^{AI}}$ — средняя длина диффузионного пробега МАП;

$\theta=\dfrac{t}{\tau^{AI}}$ — приведённое текущее время.

Используя введённые величины $l^{AI}$ и $\theta$, преобразуем уравнение диффузии (2) к виду

$$C_\theta^{AI}=l_{AI}^2 C_{xx}^{AI}-C^{AI}. \tag{9}$$

Таким образом, решение поставленной краевой задачи сводится к аналитическому решению уравнение диффузии (9) с граничными условиями (5) и начальным условием (7).

## 2. Используемый аналитический метод

Найдем решение нестационарного уравнения диффузии межузельных атомов примеси (9), используя метод функций источника или метод функций Грина [11,12]. Характерной особенностью данного уравнения является отсутствие члена, описывающего генерацию неравновесных межузельных атомов примеси. В этом случае суть метода функций Грина состоит в представлении начального распределения МАП в виде суперпозиции точечных источников, которые получаются путем умножения начального распределения на $\delta$-функцию, нахождении отклика системы на каждый источник и последующего суммирования всех откликов. Таким образом, находится отклик на произвольное начальное условие.

В работе [11] была найдена функция Грина для уравнения

$$C_\theta^{AI} = l_{AI}^2 C_{xx}^{AI} \tag{10}$$

с нулевыми граничными условиями (5) на полубесконечной прямой $[0, +\infty]$ по оси координат и с начальным условием (7).

Полученная функция Грина имеет следующий вид

$$G(x,\theta) = \frac{1}{2\sqrt{\pi l_{AI}^2 \theta}} \left\{ \exp\left[-\frac{(x-\xi)^2}{4 l_{AI}^2 \theta}\right] - \exp\left[-\frac{(x+\xi)^2}{4 l_{AI}^2 \theta}\right] \right\}. \tag{11}$$

Тогда, решение $C_1^{AI}(x,\theta)$ уравнения (10) для нулевых граничных условий имеет вид

$$C_1^{AI}(x,\theta) = \frac{1}{2\sqrt{\pi}} \int_0^{+\infty} \frac{1}{\sqrt{l_{AI}^2 \theta}} \left\{ \exp\left[-\frac{(x-\xi)^2}{4 l_{AI}^2 \theta}\right] - \exp\left[-\frac{(x+\xi)^2}{4 l_{AI}^2 \theta}\right] \right\} C_0^{AI}(\xi) d\xi. \tag{12}$$

Используем методику, описанную в [11], чтобы найти функцию Грина для уравнения диффузии (9) с граничными условиями (5).

Рассмотрим вначале с этой целью уравнение (9) на бесконечной прямой $[-\infty, +\infty]$ по оси координат и найдем его решения методом разделения переменных, представив это решение в виде произведения двух функций, одна из которых зависит только от времени $\theta$, а вторая только от пространственной координаты $x$ [11]

$$C^{AI}(x,\theta) = T(\theta)X(x).$$

Тогда

$$C^{AI}_\theta = T'X \qquad C^{AI}_{xx} = TX'' \qquad (13)$$

Подставив выражения (13) в уравнение (9), получим

$$T'X = l_{AI}^2 TX'' - TX \qquad (14)$$

Разделив правую и левую части уравнение (14) на ($l_{AI}^2 TX$), приходим к следующему уравнению

$$\frac{T'}{l_{AI}^2} = \frac{X''}{X} - \frac{1}{l_{AI}^2} = -\lambda^{*2} \qquad (15)$$

где $\lambda^{*2}$ — параметр разделения, который, как видно из Ур.(15). не зависит ни от $x$, ни от $\vartheta$, т.е. представляет собой некоторую константу.

Используя независимость параметра $\lambda^*$ от $x$ и $\theta$, получаем следующую систему уравнений:

$$\begin{aligned}\frac{T'}{T} &= -l_{AI}^2 \lambda^{*2} \\ X'' + (\lambda^{*2} - \frac{1}{l_{AI}^2})X &= 0\end{aligned} \qquad (16)$$

Решим первое уравнение системы (16):
$$\begin{aligned}\frac{T'}{T} &= -l_{AI}^2 \lambda^{*2}, \\ \frac{dT}{T} &= -l_{AI}^2 \lambda^{*2} d\theta, \\ \ln\frac{T}{T_0} &= -l_{AI}^2 \lambda^{*2}\theta, \\ T(\theta) &= T_0 \exp\left[-l_{AI}^2 \lambda^{*2}\theta\right]\end{aligned}$$

Рассмотрим теперь решение второго уравнения системы (16):

$$X'' + (\lambda^{*2} - \frac{1}{l_{AI}^2})X = 0$$

Если $\lambda^{*2} > \frac{1}{l_{AI}^2}$, то можно положить, что $\lambda^{*2} - \frac{1}{l_{AI}^2} = \lambda^2$, и второе уравнение системы(16) примет вид:

$$X'' + \lambda^2 X = 0. \tag{17}$$

Полученное уравнение (17) — типичное уравнение математической физики (например, уравнение свободных гармонических колебаний в случае, когда функция $X$ зависит от времени), которое решается следующей подстановкой $X(x) = A_0 e^{bx}$, где $A_0$ и b – некоторые постоянные.

Тогда

$$X'(x) = X'(x) = A_0 b e^{bx} \text{ и } X''(x) = A_0 b^2 e^{bx}.$$

Подставив полученные выражения в уравнение (17), сведем его к алгебраическому уравнению

$$b^2 + \lambda^2 = 0,$$
$$b^2 = -\lambda^2,$$

откуда следует, что

$$b = \pm i\sqrt{\lambda}.$$

Подставляя полученное значение $b$ в уравнение (17), найдем его решение в виде

$$X(x) = A_0 e^{\pm i\lambda x} \tag{18}$$

С учетом полученных решений для функций $T(\vartheta)$ и $X(x)$, частное решение уравнения (9) имеет вид

$$C_\lambda^{AI}(x,\theta) = A(\lambda)\exp\left[-l_{AI}^2 \lambda^{*2}\theta + i\lambda x\right] = A(\lambda)\exp\left[-l_{AI}^2(\lambda^2 + \frac{1}{l_{AI}^2})\theta + i\lambda x\right] \tag{19}$$

Здесь $\lambda$ – любое вещественное число (поэтому перед $i\lambda x$ можно взять любой знак и в данном решении мы взяли знак "+").

Общее решение исходного уравнения (9) представим в виде суперпозиции частных решений вида (19)

$$C^{AI}(x,\theta) = e^{-\theta} \int_{-\infty}^{\infty} A(\lambda) \exp\left[-l_{AI}^2 \lambda^2 \theta + i\lambda x\right] d\lambda \qquad (20)$$

Требуя выполнения начального условия при $\theta=0$, будем иметь:

$$C_0^{AI}(x) = \int_{-\infty}^{\infty} A(\lambda) \exp[i\lambda x] d\lambda, \qquad (21)$$

где $C_0^{AI}(x)$ – непрерывная функция при $\theta=0$.

Воспользуемся формулой обратного преобразования Фурье, чтобы найти функциональную зависимость $A(\lambda)$ [11]

$$A(\lambda) = \frac{1}{2\pi} \int_{-\infty}^{\infty} C_0^{AI}(\xi) \exp[i\lambda\xi] d\xi \qquad (22)$$

Подставляя (22) в (20) и меняя порядок интегрирования по $\xi$ и $\lambda$, получаем выражение для $C^{AI}(x,\theta)$

$$C^{AI}(x,\theta) = \frac{1}{2\pi} \exp\left[-l_{AI}^2 \theta\right] \int_{-\infty}^{\infty} \left[\int_{-\infty}^{\infty} C_0^{AI}(\xi) \exp[-i\lambda\xi] d\xi\right]$$

$$\times \exp\left[-l_{AI}^2 \lambda^2 \theta + i\lambda x\right] d\lambda =$$

$$= \frac{\exp\left[-l_{AI}^2 \theta\right]}{2\pi} \int_{-\infty}^{\infty} \left[\int_{-\infty}^{\infty} \exp\left[-l_{AI}^2 \lambda^2 \theta + i\lambda(x-\xi)\right] d\lambda\right] C_0^{AI}(\xi) d\xi \qquad (23)$$

Возьмём внутренний интеграл в (23)

$$\frac{1}{2\pi}\int_{-\infty}^{\infty}\exp\left[-l_{AI}^2\lambda^2\theta+i\lambda(x-\xi)\right]d\lambda = \frac{1}{2\sqrt{\pi l_{AI}^2\theta}}\exp\left[-\frac{(x-\xi)^2}{4l_{AI}^2\theta}\right] \qquad (24)$$

Подставляя (24) в (23), приходим к интегральному представлению искомого решения:

$$C^{AI}(x,\theta) = \int_{-\infty}^{\infty} G(x,\xi,\theta) C_0^{AI}(\xi) d\xi, \qquad (25)$$

где функция Грина

$$G(x,\xi,\theta) = \frac{1}{2\sqrt{\pi l_{AI}^2\theta}}\exp\left[-l_{AI}^2\theta\right]\exp\left[-\frac{(x-\xi)^2}{4l_{AI}^2\theta}\right] \qquad (26)$$

Выражения (25) и (26) представляют собой решение уравнения диффузии межузельных атомов примеси для бесконечной прямой $[-\infty, +\infty]$ по оси координат. Перейдём теперь к решению второй части задачи: получим решение на полубесконечной прямой $[0, +\infty]$ с граничными условиями (5):

$$C^{AI}(0,\theta) = C_S^{AI}, \qquad\qquad C^{AI}(+\infty,\theta) = 0, \qquad (27)$$

и начальным условием вида (6):

$$C^{AI}(x,0) = C_0^{AI}(x), \qquad (28)$$

Для того, чтобы получить единственное решение рассматриваемой краевой задачи, мы наложим условия ограниченности в бесконечности. Потребуем также в качестве дополнительного требования, чтобы искомая функция была всюду ограничена

$$C^{AI}(x,\theta) < M,$$

где М – некоторая постоянная. Отсюда следует, что и начальная функция $C^{AI}(x, 0)$ должна также удовлетворять условию ограниченности

$|C^{AI}(x, 0)| < $М.

Решение поставленной задачи можно представить в виде суммы двух функций

$$C^{AI}(x, \theta) = C_1^{AI}(x, \theta) + C_2^{AI}(x, \theta), \qquad (29)$$

где функция $C_1^{AI}(x, \theta)$ описывает влияние только начального условия при нулевых граничных условиях, а $C_2^{AI}(x, t)$ — влияние только граничных условий.

Таким образом, функции $C_1^{AI}(x, \theta)$ и $C_2^{AI}(x, \theta)$ можно определить как решения уравнения (9), удовлетворяющие условиям

$$C_1^{AI}(x, 0) = C_0^{AI}(x) \qquad C_1^{AI}(0, \theta) = 0, \qquad (30)$$

$$C_2^{AI}(x, 0) = 0. \qquad C_2^{AI}(0, \theta) = C_S^{AI} \qquad C^{AI}(+\infty; \theta) = 0. \qquad (31)$$

Очевидно, что сумма этих двух функций будет удовлетворять условиям нашей задачи (условия (5) и (6)). Действительно, функция

$$C^{AI}(x, \theta) = C_1^{AI}(x, \theta) + C_2^{AI}(x, \theta) \qquad (32)$$

удовлетворяет уравнению (9), поскольку этому уравнению удовлетворяет каждая из функций $C_1^{AI}(x, \theta)$ и $C_2^{AI}(x, \theta)$. Кроме того, непосредственная подстановка $C^{AI}(x, \theta) = C_1^{AI}(x, \theta) + C_2^{AI}(x, \theta)$ в граничные условия показывает, что эта функция удовлетворяет и заданным граничным условиям.

Введём вспомогательную функцию U(x,θ), определённую на бесконечной прямой $[-\infty, +\infty]$ и удовлетворяющую уравнению, а также условиям

U(0,θ)=0,
U(x,0)= $C_0^{AI}(x)$ для x>0.

Эту функцию, согласно лемме, доказанной в работе [11], можно определить с помощью начальной функции Ψ(x), совпадающей с $C_0^{AI}(x)$ для x>0 и являющейся нечетным продолжением $C_0^{AI}(x)$ для x<0, т.е.

$$\Psi(x) = \begin{cases} C_0^{AI}(x) & x > 0, \\ -C_0^{AI}(x) & x < 0. \end{cases}$$

Тогда

$$U(x,\theta) = \frac{1}{2\sqrt{\pi}} \int_{-\infty}^{\infty} \frac{1}{\sqrt{l_{AI}^2 \theta}} e^{-\frac{(x-\xi)^2}{4 l_{AI}^2 \theta}} \psi(\xi) d\xi$$

Рассматривая значение функции U(0,θ) только в интересующей нас области $x \geq 0$, получим:

$C^{AI}(x,\theta) = U(x,\theta)$ при $x \geq 0$.

Пользуясь определением функции Ψ(x) [11], будем иметь в результате

$$U(x,\theta) = \frac{1}{2\sqrt{\pi}} \int_{-\infty}^{0} \frac{1}{\sqrt{l_{AI}^2 \theta}} e^{-\frac{(x-\xi)^2}{4 l_{AI}^2 \theta}} \psi(\xi) d\xi + \frac{1}{2\sqrt{\pi}} \int_{0}^{\infty} \frac{1}{\sqrt{l_{AI}^2 \theta}} e^{-\frac{(x-\xi)^2}{4 l_{AI}^2 \theta}} \psi(\xi) d\xi =$$

$$= -\frac{1}{2\sqrt{\pi}} \int_{0}^{\infty} \frac{1}{\sqrt{l_{AI}^2 \theta}} e^{-\frac{(x+\xi)^2}{4 l_{AI}^2 \theta}} C_0^{AI}(\xi) d\xi + \frac{1}{2\sqrt{\pi}} \int_{0}^{\infty} \frac{1}{\sqrt{l_{AI}^2 \theta}} e^{-\frac{(x-\xi)^2}{4 l_{AI}^2 \theta}} C_0^{AI}(\xi) d\xi$$

Сводя в полученном выражении оба интеграла в один, найдем искомое решение в виде

$$C_1^{AI}(x,\theta) = \frac{1}{\sqrt{\pi}\sqrt{4 l_{AI}^2 \theta}} \int_{0}^{\infty} \left\{ \exp\left[-\frac{(x-\xi)^2}{4 l_{AI}^2 \theta}\right] - \exp\left[-\frac{(x+\xi)^2}{4 l_{AI}^2 \theta}\right] \right\} C_0^{AI}(\xi) d\xi =$$

$$= \int_{0}^{\infty} G(x,\xi,\theta) C_0^{AI}(\xi) d\xi, \qquad (33)$$

не содержащем вспомогательных функций. Заметим, что при x=0 выражение в фигурных скобках обращается в нуль и тогда $C_1^{AI}(x,\theta)=0$.

Как видно из полученного выражения, функция Грина равна

$$G(x,\xi,\theta)=\frac{1}{2\sqrt{\pi l_{AI}^2 \theta}}\left\{\exp\left[-\frac{(x-\xi)^2}{4l_{AI}^2\theta}\right]-\exp\left[-\frac{(x+\xi)^2}{4l_{AI}^2\theta}\right]\right\}. \tag{34}$$

Обратимся теперь к отысканию функции $C_2^{AI}(x,\theta)$, представляющую вторую часть решения искомой задачи с граничными условиями 1 рода и нулевым начальным условием.

Пусть

$$C_S^{AI}=const.$$

Функция

$$C_{\hom}(x,\theta)=C_S^{AI}\frac{x}{\sqrt{l_{AI}^2(\theta-\theta_0)\pi}}\int_0^\infty \exp\left(-\frac{(\xi-x)^2}{2\sqrt{l_{AI}^2(\theta-\theta_0)}}\right)$$

$$\times\frac{d\xi}{\sqrt{l_{AI}^2(\theta-\theta_0)\pi}}==C_S^{AI}\Phi\left(\frac{x}{2\sqrt{l_{AI}^2(\theta-\theta_0)}}\right) \tag{35}$$

является решением уравнения диффузии, удовлетворяющим условиям

$$C_{\hom}(x,\theta_0)=C_S^{AI} \text{ и } C_{\hom}(0,\theta)=0.$$

Входящий в выражение (35) интеграл

$$\Phi(x)=\mathrm{erf}(x)=\frac{2}{\sqrt{\pi}}\int_0^\infty \exp\left(-\frac{(\xi-x)^2}{2\sqrt{l_{AI}^2\theta}}\right)\frac{d\xi}{\sqrt{l_{AI}^2\theta\pi}}=\frac{2}{\sqrt{\pi}}\int_0^\infty \exp(-\alpha^2)d\alpha \tag{36}$$

называется интегралом ошибок (функцией ошибок). Этот интеграл часто встречается в теории вероятностей и для него существуют подробные таблицы значений. Разработан также ряд подпрограмм для вычисления функции ошибок при проведении научных и инженерных расчетов.

Тогда, функция

$$C_2^{AI}(x,\theta) = C_S^{AI} - C_{\text{hom}}(x,\theta) = C_S^{AI}\left[1 - \text{erf}(\frac{x}{2\sqrt{l_{AI}^2 \theta}})\right] \tag{37}$$

и является искомой нами функцией, так как она удовлетворяет тому же уравнению и условиям

$$C_2^{AI}(x, 0) = 0 \ (x>0) \qquad C_2^{AI}(0, \theta) = C_S^{AI}.$$

Сумма функций

$$C_1^{AI}(x, \theta) + C_2^{AI}(x, \theta)$$

дает решение рассматриваемой нами краевой задачи для полубесконечной прямой.

Таким образом, аналитическое решение уравнения диффузии (2) на полубесконечной прямой $[0, +\infty]$ имеет вид

$$C^{AI}(x,\theta) = \int_0^\infty G(x,\xi,\theta) C_0^{AI}(\xi) d\xi + C_S^{AI}\left[1 - \text{erf}(\frac{x}{2\sqrt{l_{AI}^2 \theta}})\right] \tag{38}$$

где функция Грина $G(x,\xi,\theta)$ определяется выражением (34).

Аналогичным образом можно получить решение для случая непротекания примеси через поверхность полупроводника ($x$ =0) и нулевых граничных условий в объеме полупроводника

$$\left.\frac{\partial C^{AI}}{\partial x}\right|_{x=0} = 0\ , \qquad\qquad C^{AI}(+\infty, \theta) = 0\ . \tag{39}$$

Это решение имеет вид

$$C^{AI}(x,\theta) = \int_0^\infty G(x,\xi,\theta) C_0^{AI}(\xi) d\xi \tag{40}$$

где функция Грина $G(x,\xi,\theta)$ определяется выражением

$$G(x,\xi,\theta) = \frac{1}{2\sqrt{\pi l_{AI}^2 \theta}} \left\{ \exp\left[-\frac{(x-\xi)^2}{4l_{AI}^2\theta}\right] + \exp\left[-\frac{(x+\xi)^2}{4l_{AI}^2\theta}\right] \right\}. \qquad (41)$$

## 3. Моделирование процесса диффузии межузельных атомов примеси в кремнии при низкотемпературной термообработке

Результаты моделирования процесса межузельной диффузии атомов ионно-имплантированного бора в процессе низкотемпературного термического отжига представлены на Рис.3 и Рис.4. С целью сравнения использовались экспериментальные данные [5], описанные выше. Для используемой энергии имплантации 70 кэВ значения среднего проективного пробега ионов, страгглинга среднего проективного пробега и асимметрии профиля распределения внедренных ионов равны соответственно $R_p$ = 0.219 мкм, $\Delta R_p$ = 0.0606 мкм и $Sk$ = -0.9 [13]. Для данного значения асимметрии $Sk$ положение максимума концентрации внедренных атомов $R_m$ = 0.236 мкм, то есть мало отличается от значения $R_p$. Это означает, что можно пренебречь асимметрией и описывать профиль распределения внедренных атомов после имплантации симметричным распределением Гаусса типа (7)

$$C(x, 0) = C_m \exp\left[-\frac{(x - R_p)^2}{2\Delta R_p^2}\right], \qquad (42)$$

положив значение $R_p$ равным значению $R_m$. Здесь $C_m^T$ — максимальная концентрация внедренных атомов примеси.

Для расчета значения $C_m^T$ используем формулу [13]

$$C_m^T = -\frac{Q}{\Delta R_p \sqrt{2\pi}} \times 10^{-8}, \qquad (43)$$

где доза $Q$ измеряется в ион/см², страгглинг $\Delta R_p$ в мкм и концентрация $C_m$ в мкм⁻³.

Тогда, для дозы внедренных ионов бора $10^{15}$ см$^{-2}$, которая использовалась в работе [5], величина $C_m^T$ = 6.583×10$^7$ мкм$^{-3}$. Полученное значение $C_m^T$ меньше измеренного максимального значения концентрации примеси, представленного на Рис.2, хотя на Рис.2 представлено распределение примеси после термообработки, когда первоначальное максимальное значение концентрации уменьшается в результате диффузии. Такое расхождение может быть связано либо с неточностью измерения величины дозы, либо с неточностью метода ВИМС при определении абсолютных значений. Поэтому, при расчетах было использовано значение $C_m^T$ = 9.2×10$^7$ мкм$^{-3}$, которое обеспечивает лучшее согласие с экспериментальными данными. Следует отметить, что значение $C_m^T$ соответствует максимальному значению общей концентрации примеси в начале термообработки, которая складывается из суммы концентрации примеси в положении замещения $C_m$ и концентрации межузельных атомов примеси $C_m^I$. В данной работе предполагается, что распределение межузельных атомов примеси в начальный момент термообработки также описывается распределением Гаусса и для вычисления $C_m^I$ можно использовать выражение типа (42).

На Рис.3 представлены профили распределения концентрации межузельных атомов примеси в зависимости от времени термообработки **для случая, когда на поверхности полупроводника происходит полное поглощение этих межузельных атомов** и используются граничные условия равенства нулю концентрации межузельных атомов примеси на поверхности и в объеме полупроводника. Расчеты осуществлялись с использованием аналитического выражения (38), причем символьное интегрирование осуществлялось с помощью пакета **"Mathematica 5"**. При расчетах было использовано значение $C_m^I$ = 7.0×10$^7$ мкм$^{-3}$. Как видно из Рис.3., **при увеличении времени термообработки концентрация примеси уменьшается за счёт диффузии межузельных атомов вглубь полупроводника, поглощения межузельных атомов в результате их рекомбинации с вакансиями с последующим переходом в положение замещения, а также поглощения межузельных атомов поверхностью полупроводника.**

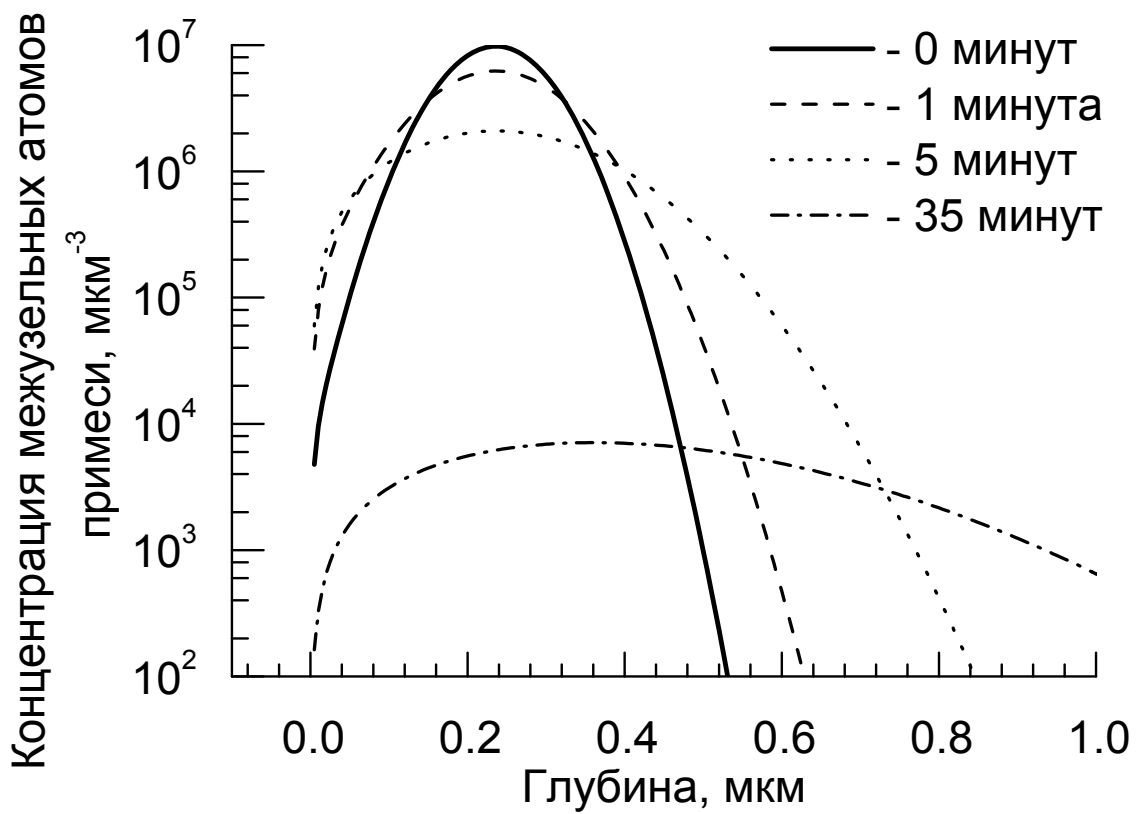

Рис.3. Профили распределения концентрации межузельных атомов примеси в зависимости от времени термообработки. Предполагается, что на поверхности происходит полное поглощение этих межузельных атомов и используется граничное условие равенства нулю концентрации межузельных атомов примеси на поверхности полупроводника

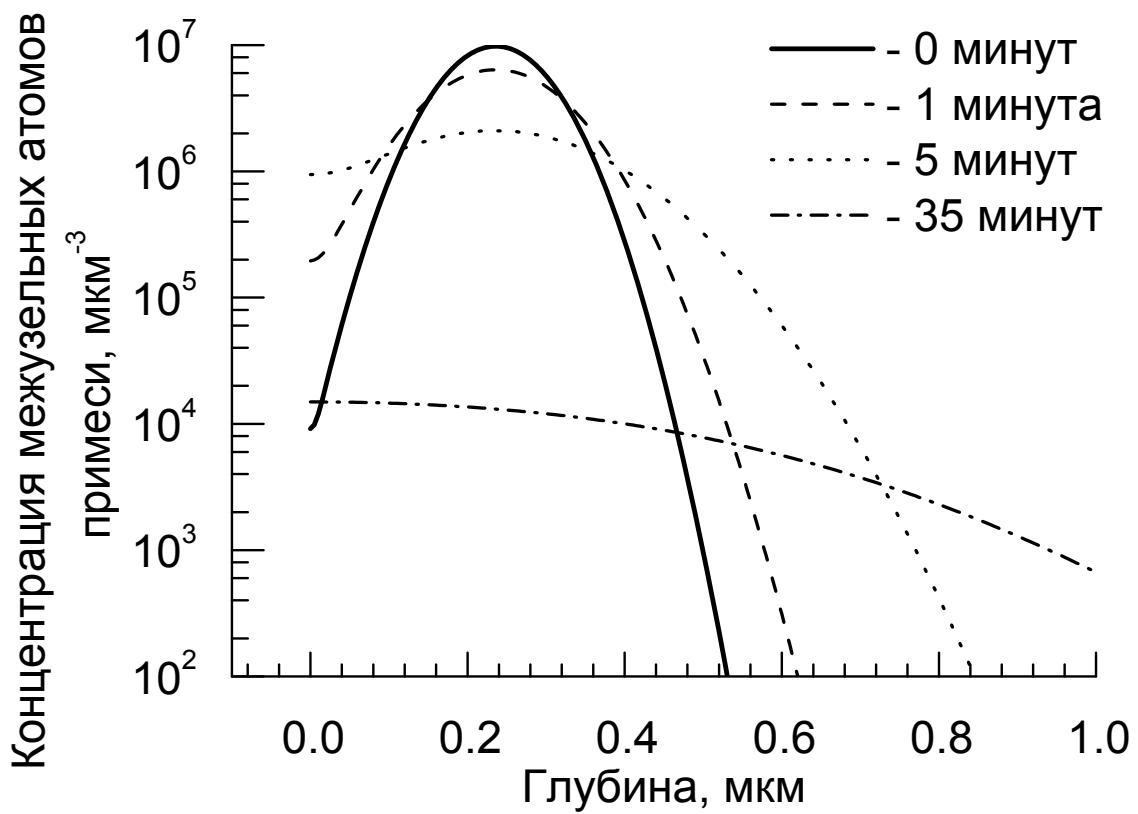

Рис.4. Профили распределения концентрации межузельных атомов примеси в зависимости от времени термообработки. Предполагается, что отсутствует поток этих межузельных атомов через поверхность полупроводника и используется граничное условие равенства нулю производной концентрации межузельных атомов примеси на поверхности полупроводника

На Рис.4 представлены аналогичные расчеты профилей распределения концентрации межузельных атомов примеси в зависимости от времени термообработки **для случая, когда отсутствует поток межузельных атомов через поверхность полупроводника и используется граничное условие равенства нулю производной концентрации межузельных атомов примеси на поверхности.** Представленные профили распределения межузельных атомов примеси были рассчитаны с использованием аналитических выражений (40) и (41). Как видно из Рис.4, **при увеличении времени диффузии концентрация примеси уменьшается только за счёт диффузии межузельных атомов вглубь полупроводника и поглощения межузельных атомов примеси в результате их рекомбинации с вакансиями.**

Рассчитанные профили распределения межузельных атомов после термообработки хорошо описывают характерные закономерности рассматриваемых диффузионных процессов, что позволяет использовать полученное аналитическое решение как для моделирования длиннопробежной миграции неравновесных межузельных атомов примеси, так и для тестирования численных решений.

# 4. Моделирование процесса перераспределения ионно-имплантированного бора в кремнии при низкотемпературной термообработке

Сравнение Рис.3 и Рис.4 с Рис.2 позволяет сделать вывод, что для моделирования процесса перераспределения ионно-имплантированного бора в кремнии при низкотемпературной термообработке лучше использовать граничное условие отсутствия потока атомов бора через поверхность полупроводника. В этом случае распределение межузельных атомов бора описывается выражениями (40) и (41).

Для нахождения общей концентрации атомов бора можно использовать уравнение (1). Интегрируя это уравнение, получаем

$$C(x,t) = \frac{1}{\tau^{AI}} \int_0^t C^{AI}(x,t)dt + C_0(x) ,\qquad(44)$$

где

$$C_0(x,0) = C_m \exp\left[-\frac{(x-R_p)^2}{2\Delta R_p^2}\right] .\qquad(45)$$

Поскольку после окончания термообработки часть атомов примеси может оставаться в межузельном положении, общую концентрацию атомов примеси получим суммированием концентраций примесных атомов в положении замещения и в межузельном положении

$$C^T(x,t) = C(x,t) + C^{AI}(x,t) .\qquad(46)$$

К сожалению, получить аналитическое выражение для интеграла (44) при $C^{AI}(x,t)$, задаваемой выражениями (40) и (41), не удается даже с помощью мощного блока символьных вычислений пакета **"Mathematica 5"**. Поэтому, нахождение значений данного интеграла осуществлялось численным методом с использование квадратурной формулы Гаусса с 16 узлами [14].

Результаты проведенных расчетов процесса перераспределения ионно-имплантированного бора в кремнии при низкотемпературной термообработке, исследованного в работе [5], представлены на Рис.5. Как видно из Рис.5., рассчитанный профиль распределения атомов бора хорошо согласуется с экспериментальными данными, включая как область максимальной концентрации атомов примеси, так и низкоконцентрационную область. При расчете были использованы следующие значения параметров, описывающих начальное распределение примесных атомов:

$C_m = 7.0 \times 10^7$ мкм$^{-3}$; $C_m^I = 7.0 \times 10^7$ мкм$^{-3}$; $R_p = 0.234$ мкм; $\Delta R_p = 0.052$ мкм,

и межузельную диффузию атомов бора:

среднее время жизни и коэффициент диффузии межузельных атомов бора $\tau^{AI} = 800$ с и $d^{AI} = 1.058 \times 10^{-5}$ мкм$^2$с$^{-1}$, соответственно; средняя длина пробега этих межузельных атомов $l_{AI} = 0.092$ мкм; длительность низкотемпературной термообработки 35 минут при 800 $^O$C.

Представленные выше значения параметров моделирования определялись из условия наилучшего согласия рассчитанного профиля распределения атомов бора с экспериментальными данными. Необходимо отметить, что определенное таким образом значение параметра $\Delta R_p$ на 14% меньше теоретического значения. Возможно, это связано с тем, что используемое нами распределение Гаусса не учитывает асимметрию профиля распределения атомов бора. В то же время значение $R_p$, определенное из экспериментальных данных практически совпадает с теоретическим значением положения максимума концентрации примесных атомов $R_m$.

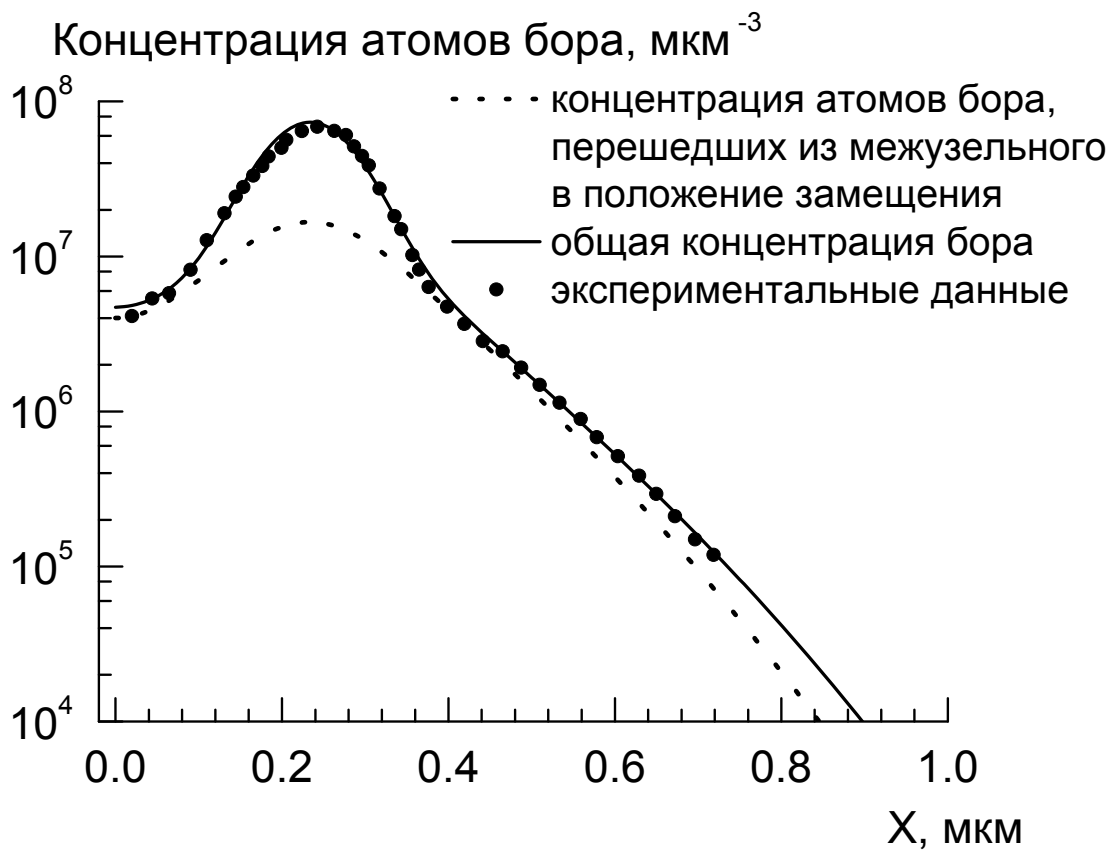

Рис.5. Рассчитанный профиль распределения концентрации атомов ионно-имплантированного бора после низкотемпературной термообработки при 800 °C в течение 35 минут.

Как следует из проведенных расчетов, разработанная модель длиннопробежной диффузии неравновесных межузельных атомов примеси позволяет добиться хорошего согласия рассчитанного профиля распределения бора после низкотемпературной обработки с экспериментальными данными [5]. Это позволяет использовать данную модель и полученные аналитические выражения для моделирования специфических профилей распределения примесных атомов при низкотемпературных отжигах ионно-имплантированных слоев. В то же время попытка объяснить образование "хвостов" в области низкой концентрации бора как результат сочетания временно ускоренной диффузии бора в ионно-имплантированных слоях с одновременно протекающими процессами образования неподвижных кластеров или преципитатов атомов бора в области высокой концентрации примеси [7,8] вряд ли будет удачной в случае низкотемпературных термообработок, поскольку, как следует из [15], концентрация атомов бора в рассматриваемом процессе [5] существенно меньше предела растворимости для данной температуры. Действительно, из Рис.2 и Рис.5 видно, что концентрация атомов бора не превышает $C_m = 8.0 \times 10^7$ мкм$^{-3}$, тогда как предел растворимости $C_{sol}$ для атомов бора при температуре 800 $^O$С согласно данным [15] равен $3.432 \times 10^8$ мкм$^{-3}$, то есть образование кластеров или преципитатов атомов бора в данном эксперименте не происходит.

# ЗАКЛЮЧЕНИЕ

**1.** Методом функций Грина получены аналитические решения нестационарного уравнения диффузии межузельных атомов примеси для случая нулевой концентрации примесных атомов на поверхности и в объеме полупроводника и для случая отсутствия потока атомов примеси через поверхность полупроводника. Предполагается, что начальное распределение примесных атомов после ионной имплантации описывается распределением Гаусса.

**2.** Для нахождения интегралов был освоен и использовался блок символьных вычислений пакета "Mathematica 5".

**3.** Как видно из проведенных расчетов, полученные аналитические решения отражают все характерные особенности рассматриваемых процессов нестационарной диффузии, а именно: диффузию межузельных атомов примеси вглубь полупроводника, поглощение межузельных атомов в результате их рекомбинации с вакансиями с последующим переходом в положение замещения, а также поглощение межузельных атомов примеси поверхностью полупроводника.

**4.** Проведено моделирование процесса перераспределения ионно-имплантированного бора при короткой низкотемпературной термообработке при 800 $^O$C. Рассчитанный профиль распределения атомов бора после термообработки хорошо согласуется с экспериментальными данными, что позволило определить ряд параметров межузельной диффузии, в частности среднюю длину пробега неравновесных межузельных атомов бора, равную 0.092 мкм при температуре 800 $^O$C.

## Практическая значимость

**1.** Полученные аналитические решения могут быть использованы для тестирования приближенных численных решений при моделировании сложных процессов межузельной диффузии.

**2.** Разработанная модель процессов межузельной диффузии и полученные аналитические выражения могут быть использованы для непосредственного моделирования процессов перераспределения ионно-имплантированных примесей при низкотемпературных термообработках, которые не описываются такими широко известными коммерческими программами моделирования технологических процессов как SUPREM и ATHENA.

# СПИСОК ИСПОЛЬЗОВАННЫХ ИСТОЧНИКОВ